\documentclass{article}
\pdfoutput=1

\usepackage{graphicx}

\usepackage[english]{babel}

\usepackage[letterpaper,top=2cm,bottom=2cm,left=3cm,right=3cm,marginparwidth=1.75cm]{geometry}

\usepackage{amsmath}
\usepackage{graphicx}
\usepackage[colorlinks=true, allcolors=blue]{hyperref}

\title{\textbf{The Hyperbolic Tangent as an Educational Tool for Teaching Variable Acceleration}}
\author{\textbf{Scott C. Scharlach$^1$}}

\begin{document}

\maketitle

\begin{center}
{\textit{$^1$Tufts University \\
574 Boston Avenue \\
Medford, MA 02155, USA}}
\end{center}

\begin{abstract}
 This paper presents an approximate analytical solution to the Falling Astronaut Problem by means of the hyperbolic tangent, and it explores the educational opportunities presented by this technique. The author's previous paper presented a function for time in terms of position $t(x)$ that modeled the motion of an astronaut as she falls from an arbitrary height to the surface of a spherical planet with no air resistance, but an exact analytical function for position in terms of time $x(t)$ was not found. This paper derives an approximate function for $x(t)$ using kinematic equations with constant acceleration and ``switch functions," specifically the hyperbolic tangent. The paper concludes with a discussion of the pedagogical implications of the technique, its potential for deepening student understanding of non-constant motion, and applications beyond the classroom.

\end{abstract}

\begin{center}
\end{center}

\section{Introduction}

Motion with non-constant acceleration appears in multitudinous areas of physics, such as the behavior of astronomical bodies, magnets, and quarks. More generally, due to the inverse-square laws of Newtonian gravity and classical electrostatics, gradients in a gravitational field or electric field are ubiquitous in nature. Indeed, constant acceleration only arises in special cases, such as a charged particle between parallel plate capacitors or an object which falls a distance that is very small compared to the size of the gravitating body. Spatially dependent acceleration is the rule, not the exception.

Despite the abundance of motion with non-constant acceleration in the universe, introductory classical mechanics courses often limit themselves to motion with constant acceleration, dismissing non-constant acceleration as being too complex for introductory students.

In this paper, the author aims to demonstrate that, for certain situations, modeling the motion of an object with non-constant acceleration is challenging but manageable for first-year undergraduate students taking an introductory physics course. To demonstrate this, the author presents a technique for approximating an analytical solution to the the Falling Astronaut Problem, a physics exercise which the author has explored in a previous paper. The paper discusses the educational opportunities that the technique presents, that being a means of conceptualizing piecewise-defined functions and time-dependent behavior in a manner unlike that which is typically discussed in introductory physics courses.

By presenting students with problems involving non-constant acceleration early in their physics curriculum, the author hopes that students will develop the understanding that non-constant acceleration is a typical scenario that physics researchers frequently encounter in their studies. In doing so, the author aims to help clarify to students what physics research will look like if and when they join a research project during their undergraduate curriculum.

\section{Problem Statement}
\label{ProblemStatement}

An astronaut is released from a height $h$ above a spherical planet with mass $M$ and a uniform radius $R$. The astronaut has negligible mass and volume compared to the planet. The astronaut falls from rest without air resistance until she reaches the surface of the planet. The situation is described by the following values:

\begin{itemize}
    \item Mass of Planet = $M$ = $5.97219 \times 10^{24}$ kg
    \item Uniform Radius of Planet = $R$ = $6.371 \times 10^{6} $ m
    \item Universal Gravitational Constant = $G$ = $6.6743 \times 10^{-11} $ m$^{3}$ kg$^{-1}$ s$^{-2}$
    \item Initial Height of Astronaut = $h$ =  $1.0000 \times 10^{7}$ m
    \item Time of Release = $t_i$ = 0 s
    \item Time of Arrival = $t_f$
\end{itemize}

\begin{figure}[ht]
\centering
\includegraphics[width=0.2\textwidth]{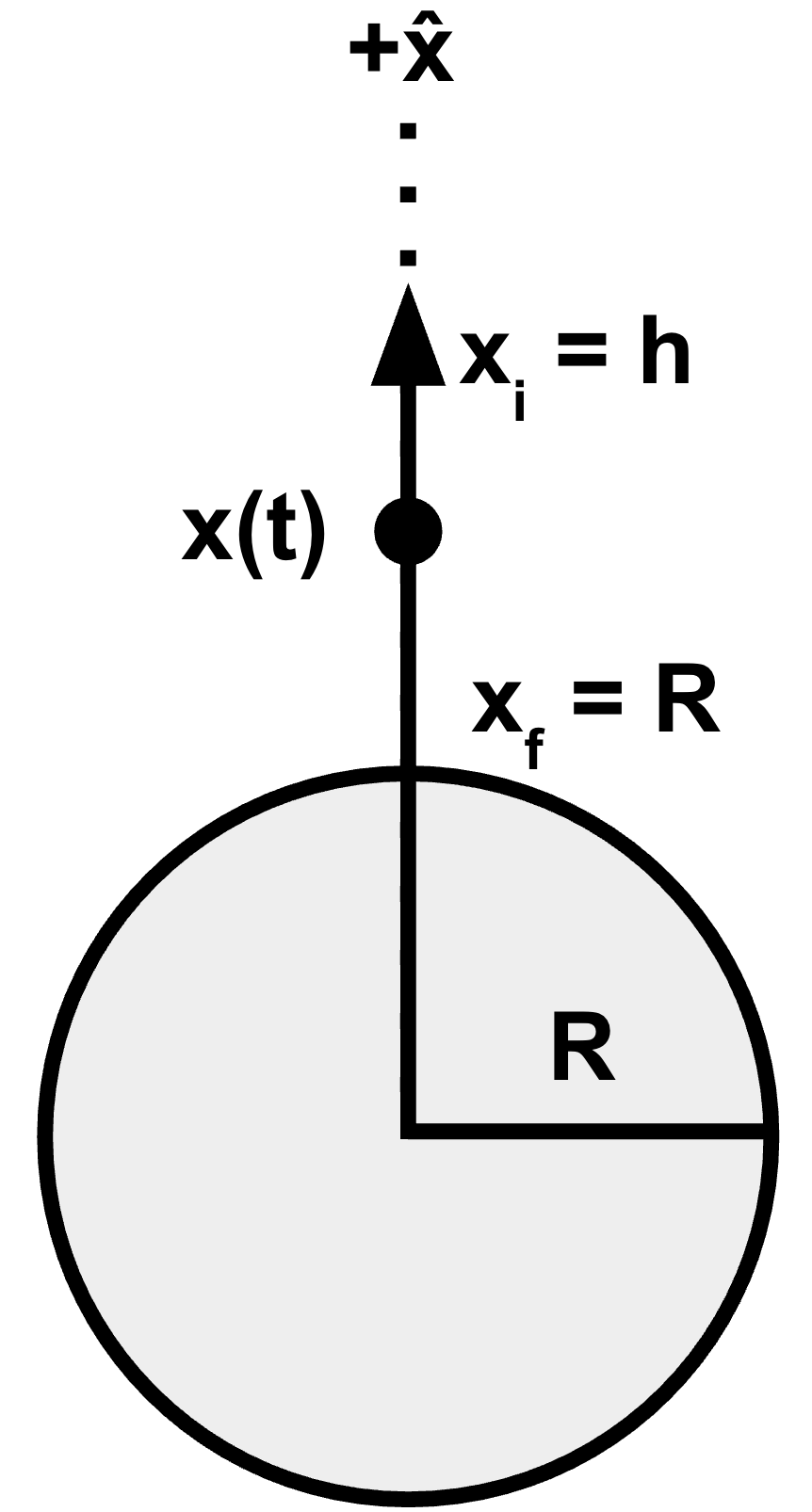}
\caption{The position of the astronaut at an arbitrary time $t$ is denoted as $x(t)$, or simply $x$. The astronaut lands and comes safely to a stop when she reaches the surface of the planet at a position $x_f = R$. The coordinate system defines positive $x$ as being directed away from the center of the planet.}
\label{diagram1}
\end{figure}

A diagram visualizing the astronaut and the planet is shown in Figure \ref{diagram1}.

Note that the acceleration due to gravity $g$ is not constant, but rather depends on the astronaut's distance from the center of the planet at any given moment.

The author's previous paper found that the motion of the astronaut obeyed the following equation:

\begin{equation}
\label{finally!}
    t(x) = \frac{h^{3/2}}{\sqrt{2GM}} \left( \frac{\pi}{2} + \sqrt{\frac{x}{h}-\frac{x^2}{h^2}} - \sin^{-1}\left(\sqrt{\frac{x}{h}}\right) \right),
\end{equation}

where $x$ is the position of the astronaut at an arbitrary time $t$.

Goal: Find a single and continuous analytical function $x(t)$ that approximates the position for any given time, as opposed to a function $t(x)$ (as in Equation \ref{finally!}) that calculates the time for a given position.

\section{Solution}
\label{Solution}

The following section is written in a pedagogical manner intended for first-year undergraduate physics students. Any deviation from a traditional academic tone is intentional for educational purposes.

\subsection{Overview}
\label{Overview}

Our goal is to find a function $x(t)$ that approximates where the astronaut is at any given time, and the function must be a closed-form equation that does not resort to infinite sums, differential equations, computational methods, or piecewise-defined functions.

A very accurate analytical approximation can be found in three steps. Firstly, we find an equation that would model the motion of the astronaut if she were to maintain her initial acceleration throughout the duration of the fall. This equation will closely model the astronaut's motion at the beginning of her fall, but as time passes, the equation will become less and less accurate.

Secondly, we will find the final velocity and acceleration of the astronaut when she arrives at the planet's surface. Using this information, we will develop an equation that would model the motion of the astronaut if she had maintained that final acceleration for the whole duration of the trip and still landed with the same final time and final velocity as with the exact solution. In other words, we imagine starting at the final moment of her fall and winding back the clock, but as we do so, we imagine that she maintains a constant acceleration. Such an equation closely models the astronaut's motion at the end of her fall, but the equation deviates greatly from the astronaut's true position for times much earlier than $t_f$.

Thirdly, we introduce a function involving the hyperbolic tangent as a way to ``switch" a function on or off. Our switch function will be approximately 0 for some input values and approximately 1 for other input values. For that reason, if a function is multiplied by the switch, the function is effectively multiplied by 0 and is ``switched off" for some input values, but it is multiplied by 1 and ``switched on" for other values.

In this way, we can form one large equation which models the motion of the astronaut at all moments in time. We can take our first equation mentioned above, which models the motion of the astronaut at early times, and multiply it by a carefully chosen switch function (based upon the hyperbolic tangent) so that it remains switched on for small values of $t$ and switched off for large values of $t$. Similarly, we can multiply the second equation by a different switch function so that it is switched off for small values of $t$ and switched on for large values of $t$. The result is  one continuous equation that models the position of the astronaut at all moments in time for the duration of the fall.

\subsection{Early-Time Kinematic Equation}

Suppose the astronaut maintains her initial acceleration for the duration of the fall. In such a scenario, the acceleration is constant, and we are justified in using one of the one-dimensional, constant-acceleration kinematic equations:

\begin{equation}
\label{kinematic}
    x_f = -\frac{1}{2}g \Delta t^2 + v_i \Delta t +x_i ,
\end{equation}

where $v_i$ is the initial velocity and $x_i$ is the initial position of the astronaut.

The astronaut falls from rest, so $v_i = 0 $ m s$^{-1}$. Also, the start of the fall is stated to occur at $t_i = 0$ s, so

\begin{equation}
    \Delta t = t_f - t_i = t_f - 0 \; \text{s} = t_f.
\end{equation}

The initial position is stated to be $x_i =h = 1.0000 \times 10^7$ m.

What is the acceleration due to gravity $g$? The value $9.8$ m s$^{-2}$ applies only near the surface of the Earth,\footnote{The problem statement describes an idealized version of Earth, one with no atmosphere and with a perfectly spherical shape. For simplicity, the paper will refer to the planet as if it is Earth.} but the astronaut begins significantly higher than the Earth's surface.

To find $g$, we will use Newton's Second Law of Motion and Newton's Law of Universal Gravitation. Newton's Second Law of Motion states that

\begin{equation}
\label{NewtonSecondLaw}
    F = m a ,
\end{equation}

where $F$ is the force applied to an object, $m$ is the mass of the object, and $a$ is the object's acceleration. If the object is accelerating because of the gravitational force, we indicate the acceleration with the symbol $g$ instead of $a$.

Newton's Universal Law of Gravity states that any two objects of masses $M$ and $m$ exert a gravitational force $F$ on each other whose magnitude depends on the distances $x$ between their centers of mass as follows:

\begin{equation}
\label{LawOfGravity}
    F = \frac{G M m}{x^2} \; .
\end{equation}

For the Falling Astronaut Problem, the force causing the astronaut to accelerate downward is the gravitational force. Thus, we can set the magnitudes of the forces provided in Equation \ref{NewtonSecondLaw} and Equation \ref{LawOfGravity} equal to each other:

\begin{equation}
    \label{acceleration}
    \begin{gathered}
        m g = \frac{ G M m}{x^2}   \\[10pt]
        \Rightarrow g = \frac{ G M}{x^2} .  \\[10pt]
    \end{gathered}
\end{equation}

The astronaut is released from rest, so $v_i = 0 $ m s$^{-1}$.

Using our values for $\Delta t$, $x_i$, $g$, and $v_i$, Equation \ref{kinematic} becomes

\begin{equation}
\label{earlykinematicnumbers}
    x_1(t) = -\frac{G M}{2 h^2}t^2 + h ,
\end{equation}

where we have added the subscript ``1" to distinguish this function from a different kinematic equation that we will find in the next section.

Let us find the numerical value for $g = \frac{G M}{h^2}$ at the initial position. Substituting our  numerical values given in the problem statement into Equation \ref{acceleration}, we calculate 

\begin{equation}
    g = \frac{ (6.6743 \times 10^{-11} \, \text{m}^3 \, \text{kg}^{-1} \, \text{s}^{-2} ) (5.97219 \times 10^{24} \, \text{kg})}{(1.0000 \times 10^7 \, \text{m})^2} = 3.9860 \, \text{m} \, \text{s}^{-2}.
\end{equation}

This is much smaller than 9.8 m s$^{-2}$, the acceleration on the surface of the Earth. This calculation demonstrates that the acceleration due to gravity is not constant along the astronaut's path, but rather there is a spatially dependent gradient for the acceleration.

Substituting our numerical values into Equation \ref{earlykinematicnumbers}, the full approximate equation becomes:

\begin{equation}
\label{kinematicearly}
    \boxed{
    x_1(t)=-\frac{1}{2}(3.9860 \, \text{m} \, \text{s}^{-2})\,t^2+(1.000\times10^7 \,\text{m}).
    }
\end{equation}

\begin{figure}[ht]
\centering
\includegraphics[width=0.70\textwidth]{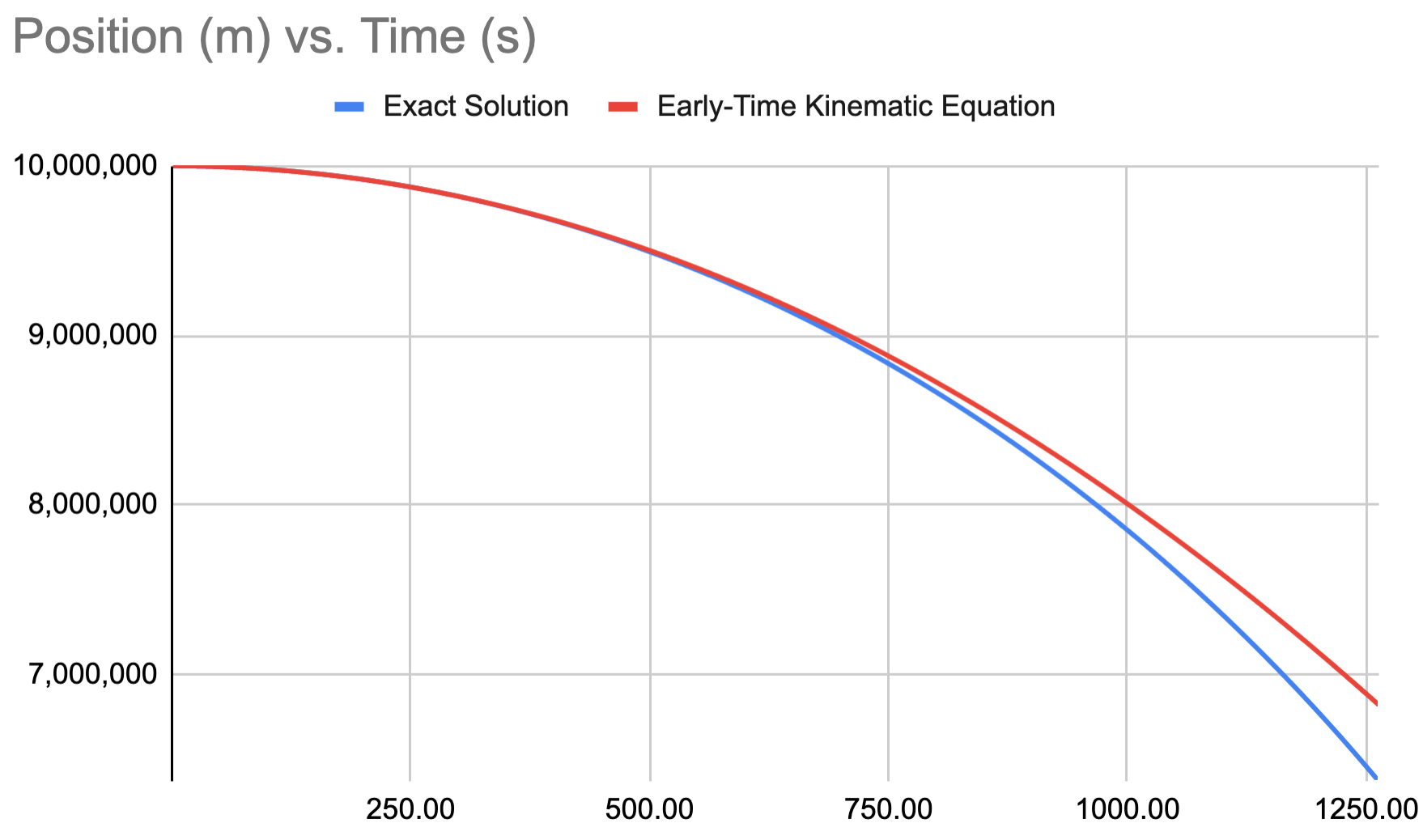}
\caption{The Early-Time Kinematic Equation accurately models the motion of the astronaut for the first half of the fall, however it underestimates astronaut's speed at later times.}
\label{EarlyKinematicEquation}
\end{figure}

\begin{figure}[ht]
\centering
\includegraphics[width=0.70\textwidth]{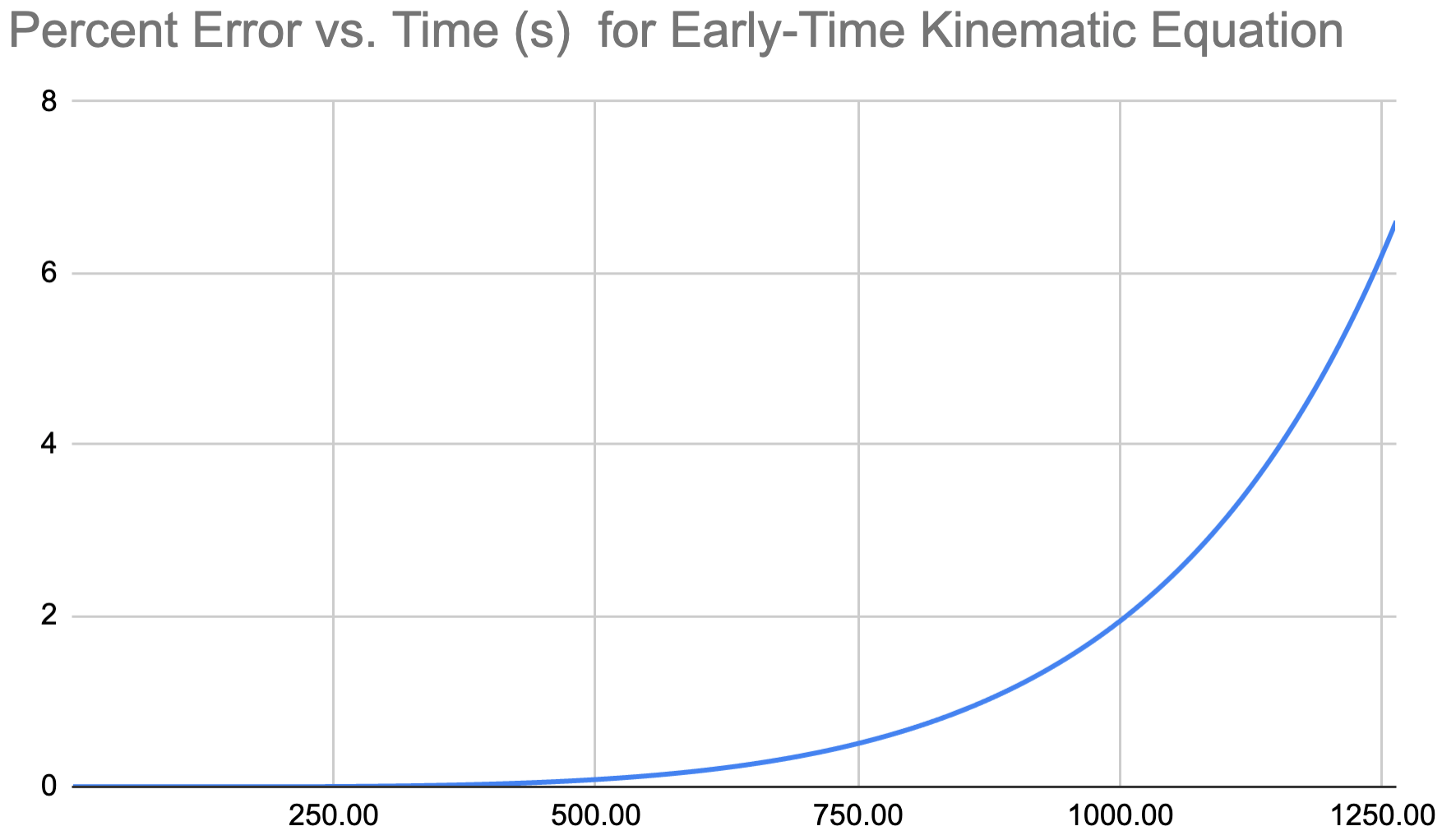}
\caption{The deviation between the exact position of the astronaut and the approximate position grows at a compounding rate as time continues, especially after 1000 s.}
\label{PercentError1}
\end{figure}

How accurate is this approximation? Equation \ref{kinematicearly} (which we will call the ``early-time kinematic equation") and Equation \ref{finally!} (the exact solution) are plotted together in Figure \ref{EarlyKinematicEquation}. The approximation remains within an error of 2$\%$ for the first 1000 seconds, as shown by Figure \ref{PercentError1}. 

Although the approximation is quite accurate for earlier times during the astronaut's fall, the margin of error increases as time progresses. The exact equation, Equation \ref{finally!}, predicts that the astronaut takes 1263 s to fall, while our approximation, Equation \ref{kinematicearly}, predicts a time of 1349 s -- an 86-second difference. Because of the inaccuracy of Equation \ref{kinematicearly} at later times, we will find another equation in the next section that models the second half of the fall.

\subsection{Late-Time Kinematic Equation}

Let us find an equation that closely models the motion for the second half of the astronaut's fall. To do this, we calculate the astronaut's final acceleration and assume that the acceleration maintained that constant value for the duration of the fall.

We are told that the final position of the astronaut is the surface of the Earth, where $x = R$. Let us substitute the numerical values given in the problem statement into Equation \ref{acceleration}:

\begin{equation}
\label{familiarg}
    g = \frac{G M}{R^2} = \frac{ (6.6743 \times 10^{-11} \, \text{m}^3 \, \text{kg}^{-1} \, \text{s}^{-2} ) (5.97219 \times 10^{24} \, \text{kg})}{(6.371 \times 10^6 \, \text{m})^2} = 9.820 \, \text{m} \, \text{s}^{-2},
\end{equation}

which matches the value 9.8 m s$^{-2}$ commonly given in physics textbooks.

It is also worth asking: what is the final velocity $v_f$ of the astronaut? Although this question may seem irrelevant, we will need to know the final velocity during a later step in the construction of our late-time kinematic equation.

To find $v_f$, let us consider conservation of energy. In the previous paper, we used conservation of energy to show that the velocity of the object at any given position is

\begin{equation}
\label{velocity}
    v(x) = - \sqrt{2 G M} \sqrt{\frac{1}{x} - \frac{1}{h}} \; .
\end{equation}

At the final position $x=R$, the velocity is

\begin{equation}
\label{finalvelocity}
    \begin{gathered}
    \resizebox{0.9\linewidth}{!}{$
    v_f = - \sqrt{(2)(6.6743 \times 10^{-11} \, \text{m}^3 \, \text{kg}^{-1} \, \text{s}^{-2})(5.97219 \times 10^{24} \, \text{kg})}\sqrt{\frac{1}{(6.371 \times 10^6 \, \text{m})} - \frac{1}{(1.000 \times 10^7 \, \text{m})}}$} \\[10pt]
    \Rightarrow v_f = -6739 \, \text{m} \, \text{s}^{-1} .
    \end{gathered}
\end{equation}

This is an extraordinary speed -- the astronaut travels over six kilometers in a single second!

We have not yet stated why it is useful to know $v_f$, but we will reveal its utility soon.

Next, let us consider the general form for our late-time kinematic equation. We are assuming that the astronaut falls at a constant acceleration of $-9.820$ m s$^{-2}$ and reaches a final velocity of $-6379$ m s$^{-1}$. We can model our new equation after the same one-dimensional constant-acceleration kinematic equation stated earlier, Equation \ref{kinematic}. Thus, our late-time kinematic equation will take the following form:

\begin{equation}
\label{kinematiclate}
    x_2(t) = -\frac{G M}{2 R^2} t^2 + v_{i2} t +x_{i2}.
\end{equation}

What is the physical meaning of $v_{i2}$ and $x_{i2}$? These values represent what the initial velocity and position would have been if the astronaut had accelerated at a constant rate of $-9.820$ m s$^{-2}$ and still arrived at the surface of the planet with a velocity of $-6739$ m s$^{-1}$ at a time of 1263 s. (As a reminder, the exact solution in Equation \ref{finally!} tells us that the astronaut reaches the planet's surface at $t$ = 1263 s.)

We want to find the numerical values of $v_{i2}$ and $x_{i2}$. Let us start with finding $v_{i2}$. We can take the time derivative of Equation \ref{kinematiclate} to acquire an equation for the velocity of the astronaut over time. This yields 

\begin{equation}
\label{velocityoft}
    v_2(t) = -\frac{G M}{R^2} t + v_{i2}.
\end{equation}

As an aside, we see that Equation \ref{velocityoft} differs from Equation \ref{velocity}, even though they both describe the velocity of the astronaut. Why is this? Firstly, Equation \ref{velocityoft} is merely an approximation of the velocity of the astronaut at late times, while Equation \ref{velocity} is exact. Secondly, Equation \ref{velocityoft} provides velocity as a function of time, while \ref{velocity} provides velocity in terms of position.

We now have a use for $v_f$. We know that $v_f$ = $-$6739 m s$^{-1}$ when $t$ = 1263.63 s. (Although the answer presented earlier was ``1263 s" with four significant figures, the actual calculated number was 1263.63 s, and thus our calculations will include the full value.) Therefore, at a time of 1263.63 s, Equation \ref{velocityoft} has only one unknown value, $v_{i2}$. We thus have enough information to calculate $v_{i2}$.

To determine the numerical value of $v_{i2}$, let us algebraically isolate $v_{i2}$ then substitute our numerical quantities into Equation \ref{velocityoft}:

\begin{equation}
\label{velocityoft2}
    \begin{gathered}
    \Rightarrow v_{i2} = v_2(t) + \frac{G M}{R^2} t \\[10pt]
    \Rightarrow v_{i2} = -6739 \, \text{m} \, \text{s}^{-1}+ (9.820 \, \text{m} \, \text{s}^{-2} ) (1263.63 \text{s}) \\[10pt]
    \Rightarrow v_{i2} = 5669 \, \text{m} \, \text{s}^{-1} .
    \end{gathered}
\end{equation}

The physical interpretation of this equation is as follows: if the astronaut had accelerated at a constant rate $-9.820$ m s$^{-2}$ and arrived at her destination at $t$ = 1263 s with a final velocity of $-6739$ m s$^{-1}$, then she would have begun her journey with an upward velocity of 5669 m s$^{-1}$.

Let us now find $x_i$. We know from the exact solution that $t=1263.63$ s when $x=R=6.371 \times10^6\,$m. We found the numerical value for $\frac{G M}{R^2}$ in Equation \ref{familiarg} and the numerical value for $v_{i2}$ in Equation \ref{velocityoft2}.

Using Equation \ref{kinematiclate}, we can isolate $x_i$ and substitute in our numerical values:

\begin{equation}
\label{xi2}
    \begin{gathered}
         \Rightarrow x_{i2} =x_2(t) + \frac{1}{2} \frac{G M}{R^2} \, t^2 - v_{i2} t \\[10pt]
        \Rightarrow x_{i2} =6.371 \times10^6\,\text{m} +\frac{1}{2}(9.820 \, \text{m} \, \text{s}^{-2})(1263.63 \text{s})^2 - (5669 \, \text{m} \, \text{s}^{-1}) (1263.63 \text{s}) \\[10pt]
        \Rightarrow x_{i2} = 7.052 \times 10^6 \, \text{m}. 
    \end{gathered}
\end{equation}

This equation tells us that, in this constant-acceleration scenario, the astronaut would have been launched from a height of 7.052 $\times$ 10$^6$ m.

Finally, substituting all of our numerical values into Equation \ref{kinematiclate}, we have arrived at the late-time kinematic equation for the approximate position of the astronaut:

\begin{equation}
\label{numericalkinematiclate}
    \boxed{
    x_2(t) = -\frac{1}{2}(9.820 \, \text{m} \, \text{s}^{-2}) t^2 + (5669 \, \text{m} \, \text{s}^{-1}) t +7.052 \times 10^6 \, \text{m}.
    }
\end{equation}

\begin{figure}[ht]
\centering
\includegraphics[width=0.70\textwidth]{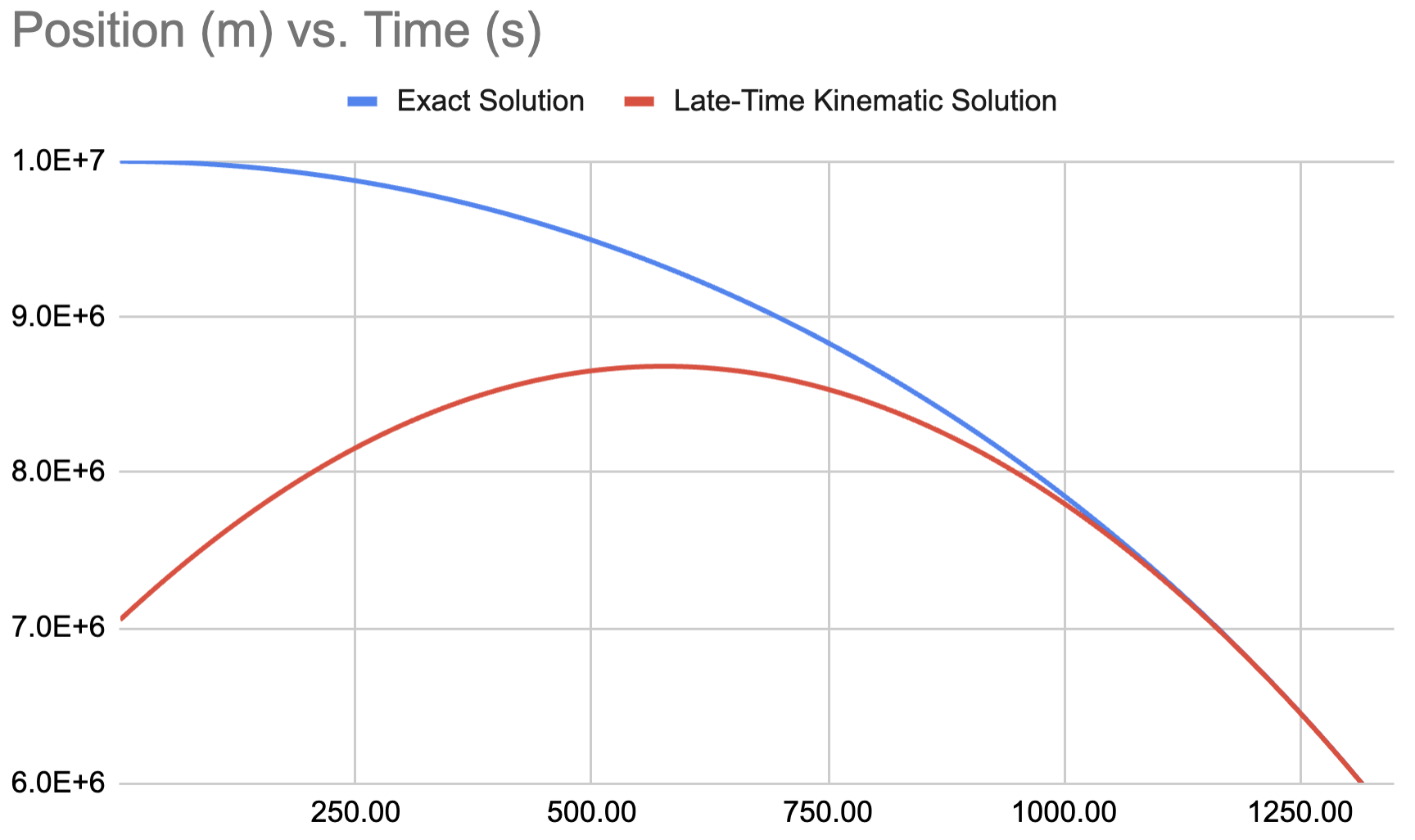}
\caption{In this constant-acceleration model, the astronaut would have been launched upward with a large initial velocity and then fallen back down. Such a model is unhelpful for representing the true motion of the astronaut at earlier times, but the approximation is quite accurate for the final several hundred seconds of the fall.}
\label{figLateTimeGraph}
\end{figure}

Figure \ref{figLateTimeGraph} compares Equation \ref{numericalkinematiclate}, the late-time kinematic equation, to the exact solution. We see that the approximation is very precise for times after approximately 900 s, but the deviation compounds as time decreases, leading to a graph with almost no resemblance at the earliest moments of the fall.

\begin{figure}[ht]
\centering
\includegraphics[width=0.70\textwidth]{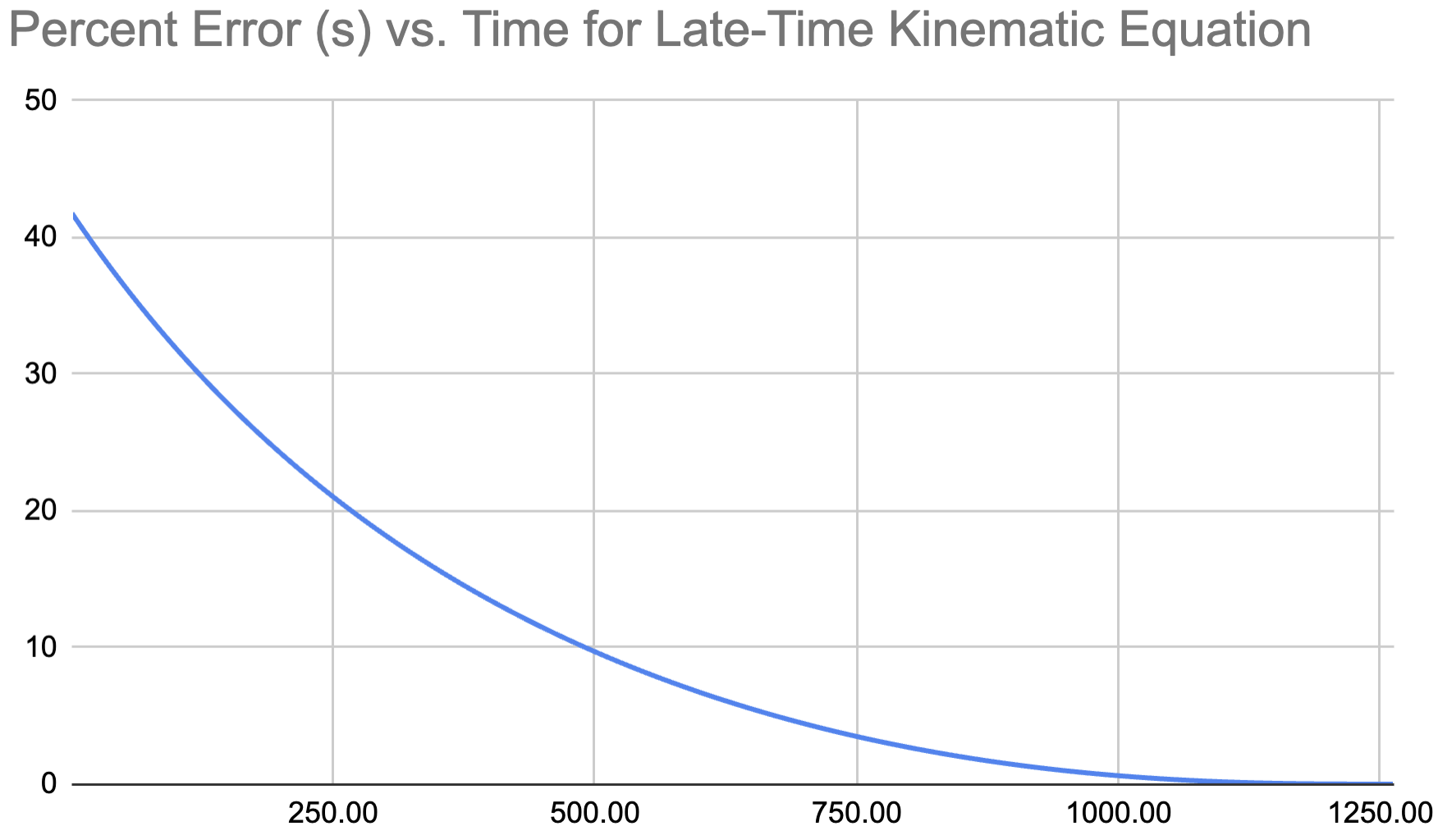}
\caption{The late-time kinematic equation is primarily useful for times later than 900 s, when the percent error is less than 1.5\%.}
\label{figLateTimePercent}
\end{figure}

Figure \ref{figLateTimePercent} displays the percent error between the late-time kinematic equation (Equation \ref{numericalkinematiclate}) and the exact solution over time. The percent error remains under five percent for times as early as $t = 674$ s.

We now have two equations, one which approximates the position of the astronaut at earlier times in the fall (Equation \ref{kinematicearly}) and one which approximates the position at later times (Equation \ref{numericalkinematiclate}). In the next section, we will discuss methods for interweaving these two equations into one continuous equation that describes the position of the astronaut for the duration of the entire fall.

\subsection{Piecewise-Defined Function}

We currently have two equations that describe the motion of the astronaut, but each equation is only accurate for a specific interval of time. The goal of this section is to combine our two kinematic equations so that each equation only has a significant impact on the final result within its corresponding time interval. In other words, we seek an equation that smoothly transitions from $x_1(t)$ to $x_2(t)$.

The most straightforward manner of doing this is to define a piecewise defined function, a function that exhibits different behaviors within different intervals.

Let us construct a piecewise defined function for the motion of the astronaut. At $t=915 \,$s, both $x_1(t)$ and $x_2(t)$ deviate from the exact solution by approximately 106,000 m. This moment in time serves as a natural place to change the behavior of the function, as it marks when $x_2(t)$ becomes accurate than $x_1(t)$, at least in terms of the exact difference rather than the percent error. Our piecewise-defined function is thus:

\begin{equation}
\label{piecewiseequ}
x(t)=
\begin{cases} 
        -\frac{1}{2}(3.9860 \, \text{m} \, \text{s}^{-2})\,t^2+(1.000\times10^7 \,\text{m}) & t \leq 915 s \\
        -\frac{1}{2}(9.820 \, \text{m} \, \text{s}^{-2}) t^2 + (5669 \, \text{m} \, \text{s}^{-1}) t +7.052 \times 10^6 \, \text{m} & t  > 915 s 
   \end{cases}
    \; .
\end{equation}

\begin{figure}[ht]
\centering
\includegraphics[width=0.70\textwidth]{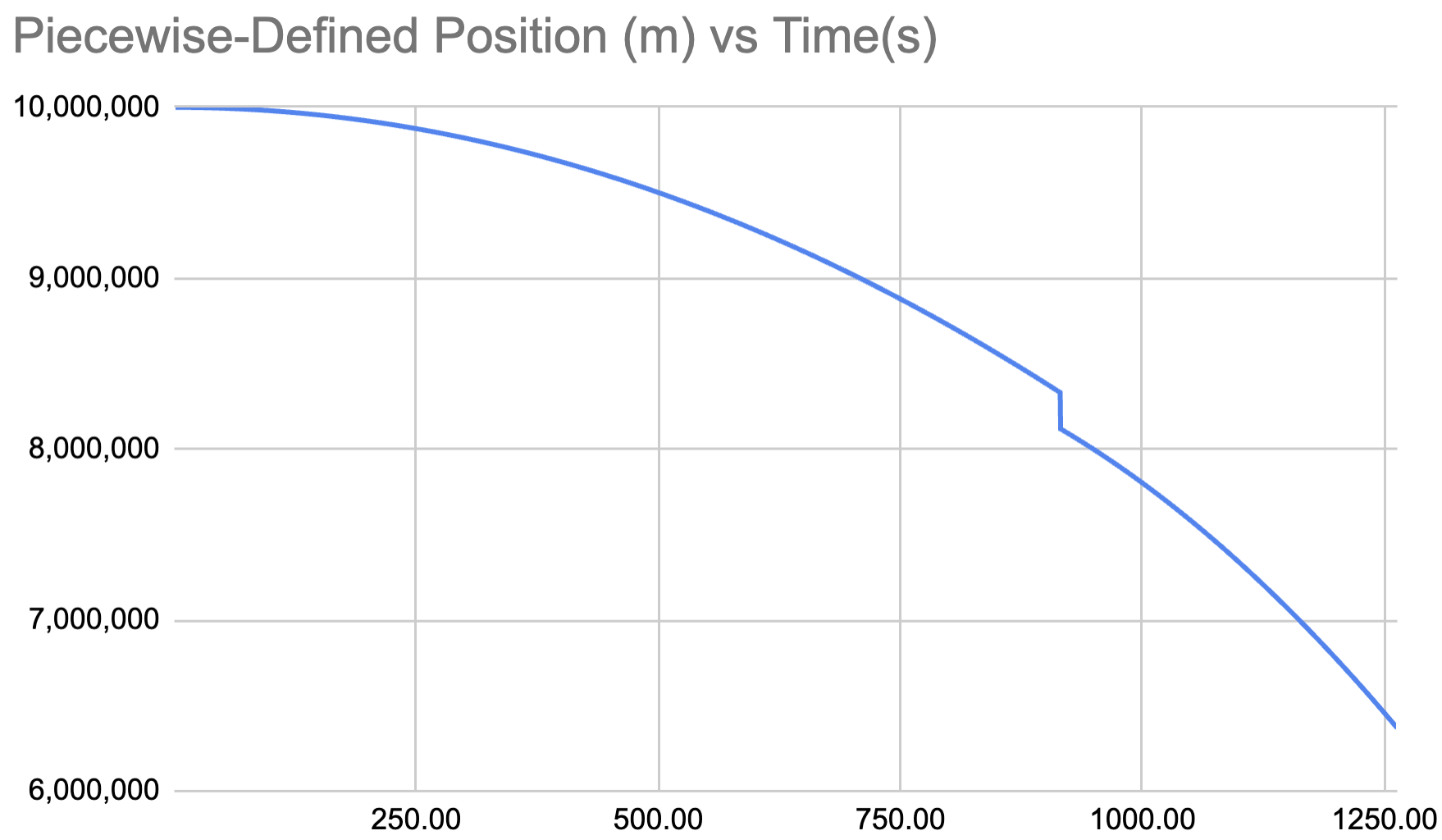}
\caption{Although the piecewise-defined function retains a low percent error, it exhibits a discontinuity at t = 915 s which differs greatly from the true behavior of the astronaut.}
\label{Piecewise}
\end{figure}

Figure \ref{Piecewise} graphs our piecewise-defined function (Equation \ref{piecewiseequ}) over time. The percent error never exceeds 1.4$\%$.

Although Equation \ref{piecewiseequ} is a single equation that models the motion of the astronaut for the whole duration of the flight, it possesses certain drawbacks. Firstly, it violates ``time translation symmetry." All of the fundamental laws of physics obey time translation symmetry, which means that the same core equations hold true at all moments in time. In other words, the laws of physics are the same now as they were at the earliest moments of the universe. Time translation symmetry is of crucial importance in physics, as it gives rise to conservation of energy via Noether's theorem. Furthermore, it is one of the fundamental symmetries in the Poincaré group, which applies to every branch of physics. For this reason, it is pedagogically imperative for instructors to help students conceptualize time translation symmetry as an inseparable feature of the universe. Equation \ref{piecewiseequ}, in contrast, abruptly switches from one behavior to another. Thus, our piecewise-defined function is pedagogically counterproductive.

A second drawback is that the function exhibits a stark discontinuity at $t=915$ s. The equation implies that the astronaut effectively teleports in an instant from $x = 8.33141 \times 10^6$ m to $x=8.12836 \times 10^6$ m, a distance of over 200,000 m. Such an implication clearly contradicts the actual behavior of the astronaut.

In the next section, we will introduce the hyperbolic tangent as a means of ensuring time translation symmetry and continuity for our equation of motion.

\subsection{The Hyperbolic Tangent}

As mentioned in Section \ref{Overview}: Overview, we can write our two kinematic equations as one continuous equation using a ``switch." A mathematical switch is a function which is approximately 1 for some values of the input variable, such as 0 s $< t < 915$ s, and approximately 0 for all other values of the input variable, such as 915 s $\le t < 1263$ s.

Our goal is to rewrite our equation of motion so that it takes the following form:

\begin{equation}
\label{earlyswitch}
    x(t) = x_1(t) \times \text{switch}_1(t)+x_2(t) \times \text{switch}_2(t),
\end{equation}

where ``switch$_1(t)$" and ``switch$_2(t)$" represent the switch functions that we will determine in this section.

A natural starting place for determining the appropriate switch functions is the hyperbolic tangent, written as ``$\tanh(x)$."\footnote{When reading aloud an equation with a hyperbolic tangent, the symbol ``tanh" is sometimes pronounced ``tansh."} It is is defined by

\begin{equation}
    \tanh(x) = \frac{e^x-e^{-x}}{e^x+e^{-x}}.
\end{equation}

\begin{figure}[ht]
\centering
\includegraphics[width=0.70\textwidth]{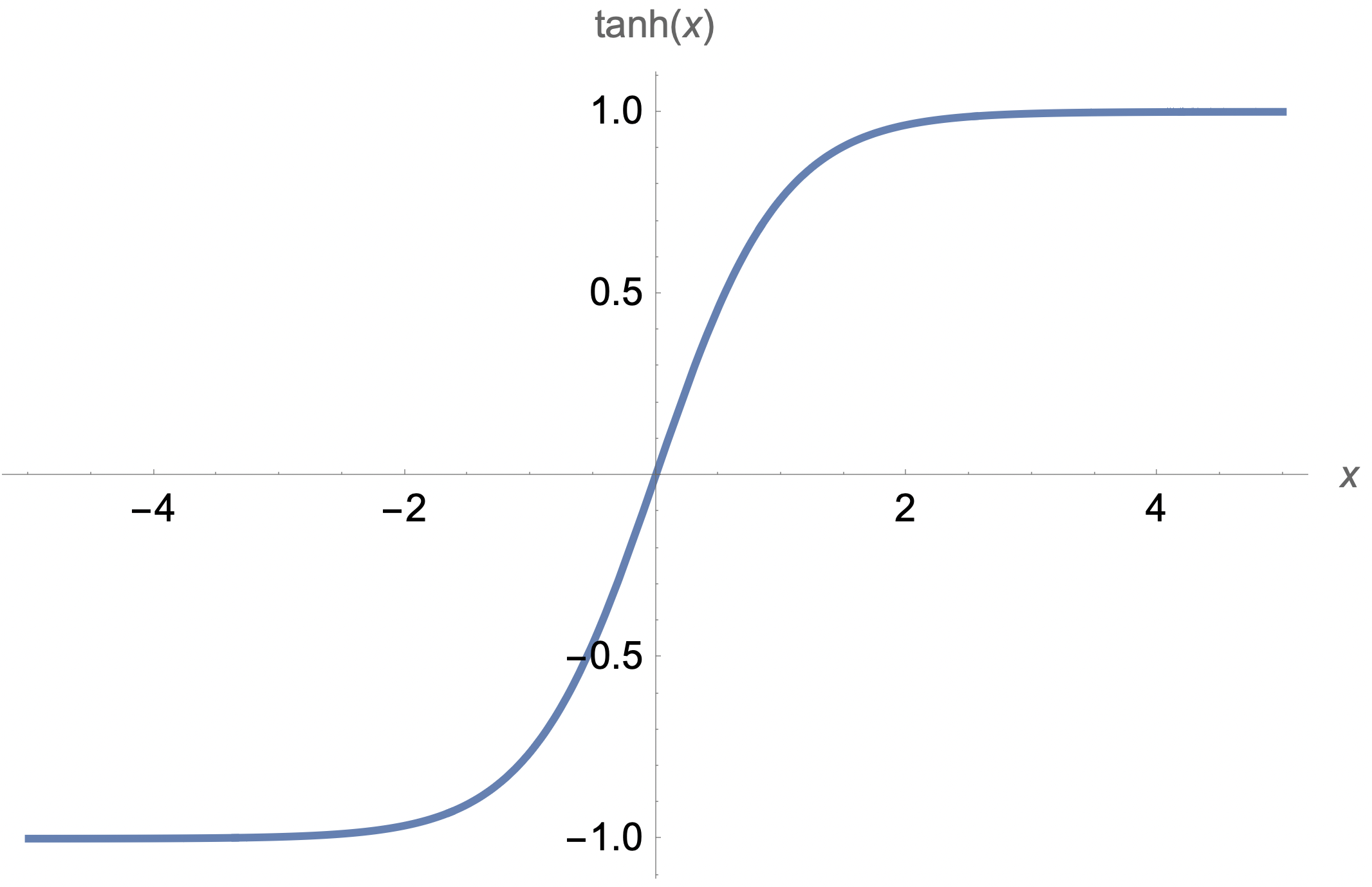}
\caption{The hyperbolic tangent is nearly $-$1 for negative values of x, but it ``switches" to nearly 1 for positive values of x. It is an example of an ``s-curve" in mathematics, so named for the S-like shape of the graphed function.}
\label{Tanh}
\end{figure}

Figure \ref{Tanh} plots the hyperbolic tangent. This function is close to what we desire: it is nearly $-$1 for some values and nearly 1 for others, and the function smoothly transitions from $-$1 to 1 without discontinuity.

However, the hyperbolic tangent needs modifications if we are to use it as our switch function. For example, we want a function that ranges from 0 to 1, not $-$1 to 1. We can raise the function up by 1 unit by writing the function as ``$1+\tanh(x)$," causing the function to range from 0 to 2. We can then change the upper limit by dividing the whole term by 2. Thus, our function become

\begin{equation}
\label{modifiedtanh}
    f(x) = \frac{1+\tanh(x)}{2}.
\end{equation}

\begin{figure}[ht]
\centering
\includegraphics[width=0.70\textwidth]{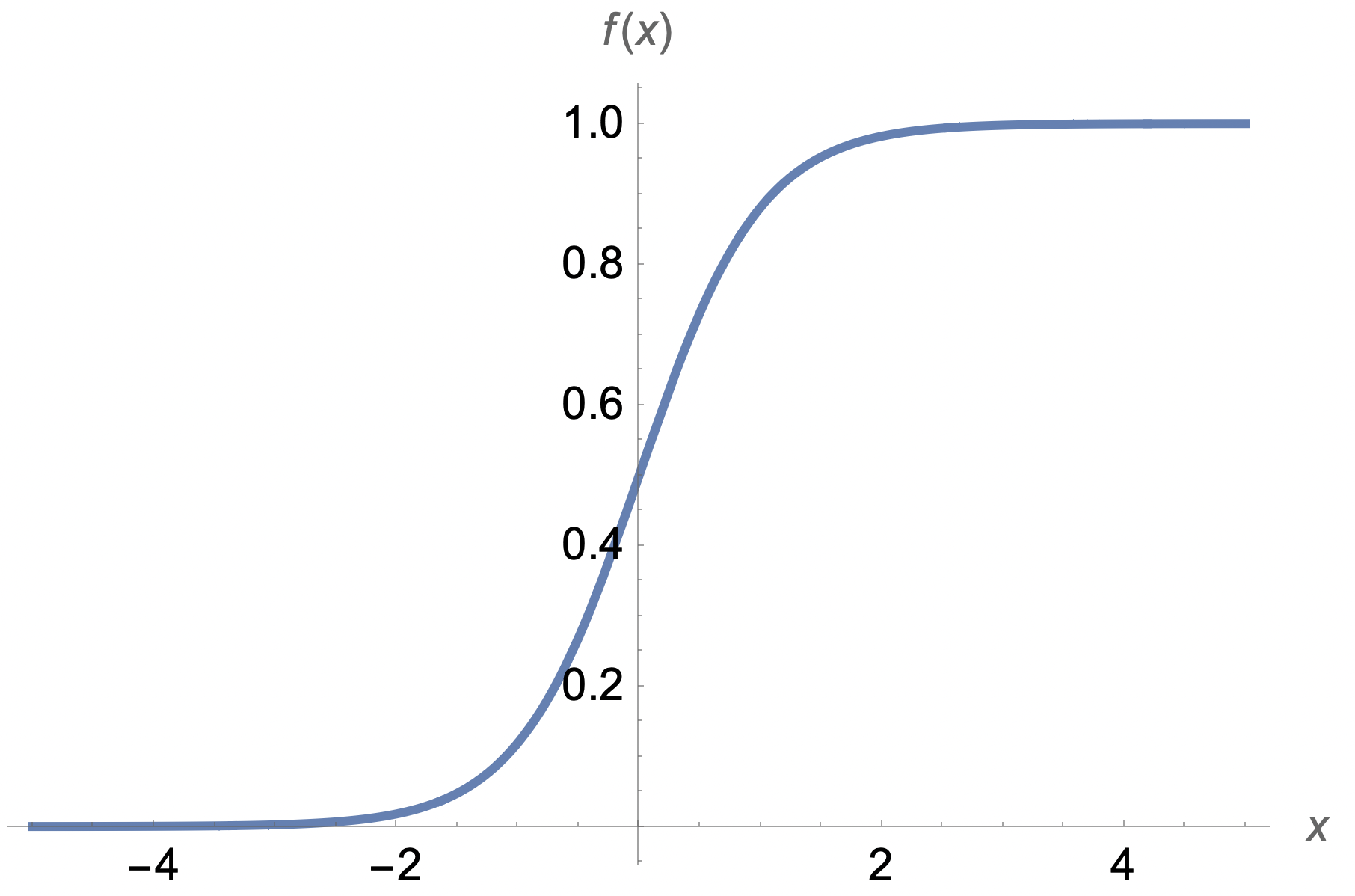}
\caption{The modified hyperbolic tangent now ranges from 0 to 1. When multiplied by another function, say $y(x)$, the function $y(x)$ is ``switched off" for negative values of $x$ and ``switched on" for positive values of $x$.}
\label{BetterTanh}
\end{figure}

Figure \ref{BetterTanh} plots our improved hyperbolic tangent function. This is precisely what we desire -- it is nearly 0 at some values and nearly 1 at others, and it smoothly transitions from nearly 0 to nearly 1. We have successfully taken the next step toward finding our switch functions.

As an aside, it is important to note that there is at least one other equation that could also serve as our switch function: the sigmoid, denoted as $\sigma(x)$. It is defined by 

\begin{equation}
    \sigma(x) = \frac{1}{1+e^{-x}}.
\end{equation}

The sigmoid produces an ``s-curve" similar to our modified hyperbolic tangent $f(x)$. Indeed, both the modified hyperbolic tangent and the sigmoid are appropriate for this exercise, but this paper will use only the modified hyperbolic tangent for simplicity and consistency.

Our modified hyperbolic tangent in Equation \ref{modifiedtanh} is still incomplete. For example, we want the switch to occur at $t=915$ s, not  $t=0$ s. To improve our switch function, let us rewrite Equation \ref{modifiedtanh} in a more general way, one which has constants that we are free to change. We will name our generalized function $s(t)$:

\begin{equation}
\label{modifiedtanh2}
    s(t) = \frac{1+c \, \tanh\big(\alpha(t-t_c) \big)}{2}.
\end{equation}

Let us change these constants one at a time and explore the constants' effects on the function. Students are encouraged to follow along by opening a link to the Desmos graphing calculator, which allows students to modify constants and visualize their effects in real time.\footnote{\raggedright\hangindent=1em\hangafter=1
Weblink to Desmos graphing calculator for Equation~\ref{modifiedtanh2}: \texttt{https://www.desmos.com/calculator/6f2j3n7iti}. Credit: Desmos Studio, Public Benefit Corporation.
}

Upon changing $t_c$, we see that increases in $t_c$ shift the graph of $s(t)$ to the right. Thus, our choice of $t_c$ allows us to choose when the switch function transitions from 0 to 1. In other words, $t_c$ determines when the switch is ``flipped." In the default form of the hyperbolic tangent function, the switch is flipped at $t=$ 0 s. However, for the Falling Astronaut Problem, we have established that we want our switches to be flipped at $t=915$ s. We will therefore choose $t_c = 915$ s for both switch functions in Equation \ref{earlyswitch}.

Let us now consider the second constant, $c$. When we change $c$ from 1 to $-$1, we see that the function is effectively reflected across the horizontal line $s=1$. In other words, instead of beginning near 0 then increasing to nearly 1, the function begins near 1 then decreases to nearly 0. In the context of our kinematic equations, the early-time kinematic equation (Equation \ref{kinematicearly}) should be ``on" at the beginning of the time interval and become switched off at $t=915$ s. For that reason, we want $c =-1$.

In contrast, the late-time kinematic equation (Equation \ref{numericalkinematiclate}) is ``off" at the beginning of the astronaut's fall and is switched on at $t=915$ s. Its corresponding switch function should thus have $c=+1$. In choosing our values for $c$, we effectively switch off Equation \ref{kinematicearly} at the same moment that we switch on Equation \ref{numericalkinematiclate}.

The final constant is $\alpha$. Increasing $\alpha$ to values greater than 1 ``tightens" the graph, making the transition from 0 to 1 more abrupt. When decreasing $\alpha$ to values $0 <\alpha<1$, the transition occurs more slowly. A value of $\alpha =0$ corresponds to a constant value of 0.5 without any switch-flip.

We have not yet discovered what the optimal value for $\alpha$ is. For now, let us keep $\alpha = 1$, and we will return to the topic of what $\alpha$ ought to be in the next section.

We can now construct a single continuous kinematic equation for the motion of the astronaut. Let us use Equation \ref{modifiedtanh2} for the structure of our two switch functions, and let us use the values for $c$, $t_c$, and $\alpha$ described above. When we insert these values into Equation \ref{earlyswitch}, we acquire

\begin{equation}
\label{betterswitch}
    x(t) = x_1(t) \left( \frac{1-\tanh(t-915 \, \text{s})}{2} \right) +x_2(t) \left( \frac{1+\tanh(t-915 \, \text{s})}{2} \right),
\end{equation}

where $x_1(t)$ and $x_2(t)$ are Equation \ref{kinematicearly} and Equation \ref{numericalkinematiclate}, respectively.

\begin{figure}[ht]
\centering
\includegraphics[width=0.70\textwidth]{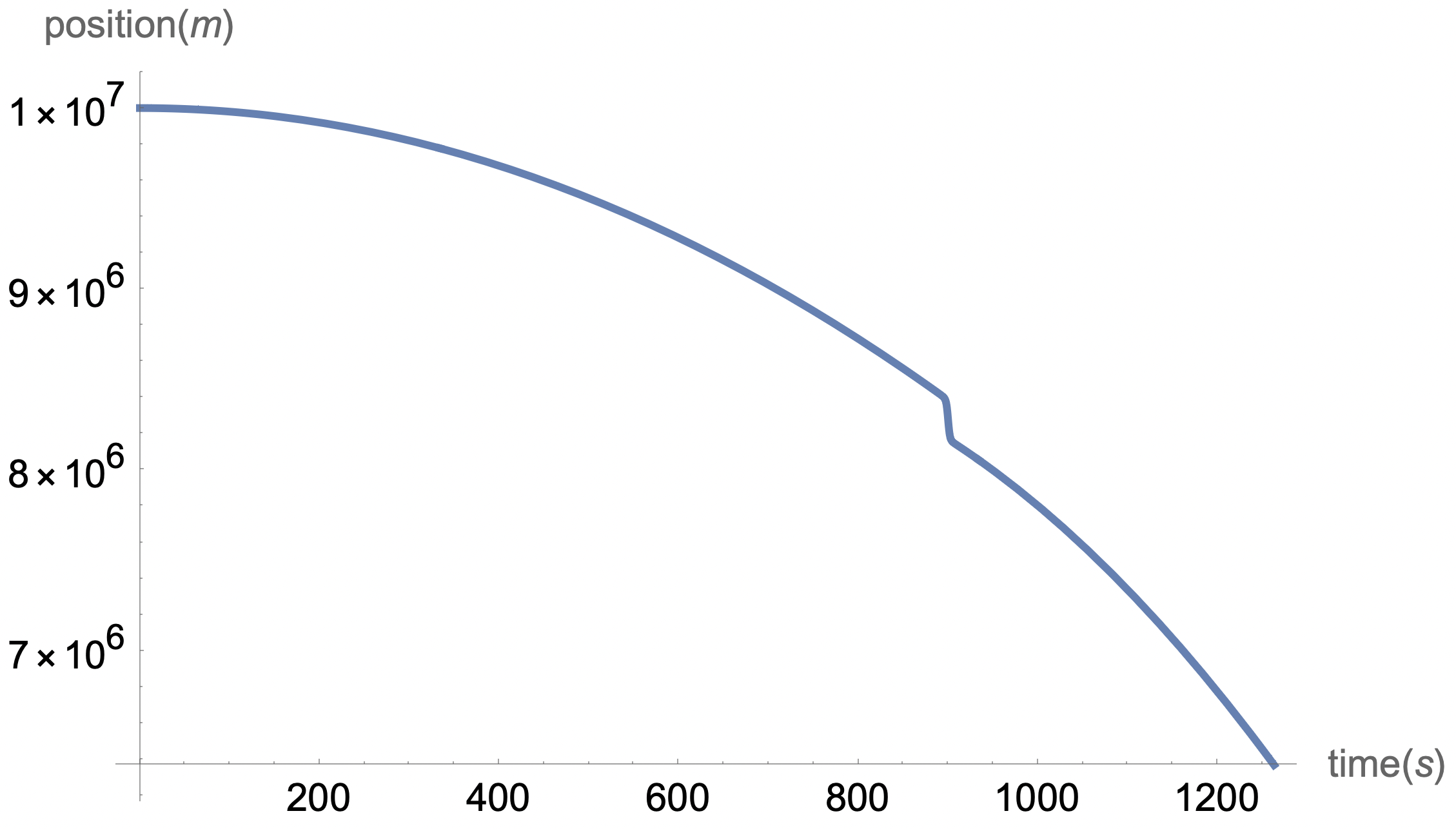}
\caption{The apparent discontinuity at $t=915$ s is actually a very quick but smooth and continuous transition. Equation \ref{betterswitch} maintains time translation symmetry.}
\label{BetterSwitch}
\end{figure}

Figure \ref{BetterSwitch} graphs Equation \ref{betterswitch}. Initially, there may appear to be no improvement over the piecewise-defined function displayed in Figure \ref{Piecewise}, as there is still a moment in time in which the astronaut abruptly drops a great distance. However, there are two benefits to Equation \ref{betterswitch}.

Firstly, the function is continuous. If we were to zoom into Figure \ref{Piecewise}, we would see a near-vertical line at $t=915$ s, which corresponds to the astronaut effectively teleporting from one position to another. In contrast, if we were to zoom into Figure \ref{BetterSwitch}, we would see a quick but nonetheless smooth transition from one position to another. The equation implies the astronaut suddenly speeds up and drops a great distance over a short amount of time, then suddenly slows back down, which admittedly is implausible. But it does not imply that the astronaut teleported. Although the function is imperfect, it preserves time translation symmetry and is continuous, which is an improvement over Equation \ref{piecewiseequ}. In other words, we have improved our equation of motion from a pedagogical perspective, even if it is no more mathematically accurate.

The second benefit to Equation \ref{betterswitch} lies with our ability to change $\alpha$. As mentioned above, decreasing alpha to a value less than 1 but greater than 0 causes the switch functions to be more gradual in their transition from 0 to 1. By changing $\alpha$ to a small value such as 0.1, the switch functions are less analogous to an on-off switch and more analogous to a dial. In the next section, we will see that choosing an appropriate value for $\alpha$ allows us to slowly ``dial down" $x_1(t)$ at the same rate that we ``dial up" $x_1(t)$, which will result in a much more smooth and accurate equation of motion. 

\subsection{\texorpdfstring{Determining the Rate of Transition, $\alpha$}{Determining the Rate of Transition, alpha}}

As mentioned in the previous section, the constant $\alpha$ as shown in Equation \ref{modifiedtanh2} tells us how quickly the switch function transitions from 0 to 1. We assumed $\alpha$ to be 1 in Equation \ref{betterswitch}, but there is still an abrupt change in position around $t=915$ s, implying that 1 is not the most accurate value for $\alpha.$ Let us graph our function in Desmos, but the only constant we are free to change is $\alpha$.\footnote{Weblink: $\texttt{https://www.desmos.com/calculator/jsxy1vso0i.}$ Credit: Ibid.} We see that a smaller value for $\alpha$ such as 0.1 provides a much smoother transition between $x_1(t)$ and $x_2(t)$, while an $\alpha$ of 0 deviates greatly from the exact solution. Thus, $\alpha$ must be between 0 and 0.1.

When we adjust the slider, we see that the ideal value for $\alpha$ is around 0.006. For values of $\alpha$ larger than 0.006, the region near $t=915$ s looks slightly warped. For values smaller than 0.006, the initial position ceases to be approximately 1.0000 $\times$ 10$^7$ m.

How can we determine the ideal value for $\alpha$? To answer this question, we can use a computer program using Python. The computer checks values of $\alpha$ from 0 to 1 at intervals of 0.0001. For each value of $\alpha$, the computer compares our approximate equation to the exact solution.\footnote{To be precise, the approximate solution is compared to a numerical approximation found in the previous paper; the accuracy is high enough that we can treat the numerical approximation data as if it is exact.} For each function, the computer runs a ``Least Squares Fit," which measures how closely a function models a dataset. Once the program has run this test, it mathematically calculates a number called the ``$R^2$" value, which ranges from 0 to 1. An $R^2$ of 1 means that our function matches the data with absolute perfection, while an $R^2$ of 0 means there is no similarity between the data and the function at all. Therefore, the larger the $R^2$, the better the fit. We will program our Python code to tell us which $\alpha$ corresponds to the highest $R^2$ value.

The Python code can be found in the form of a Jupyter Notebook titled \path{TangentApproximation.ipynb}, attached in the files of the arXiv submission for this paper. It reads the dataset titled \path{NumericalData.csv}, which is also included in the arxiv submission.

\begin{figure}[ht]
\centering
\includegraphics[width=0.7\textwidth]{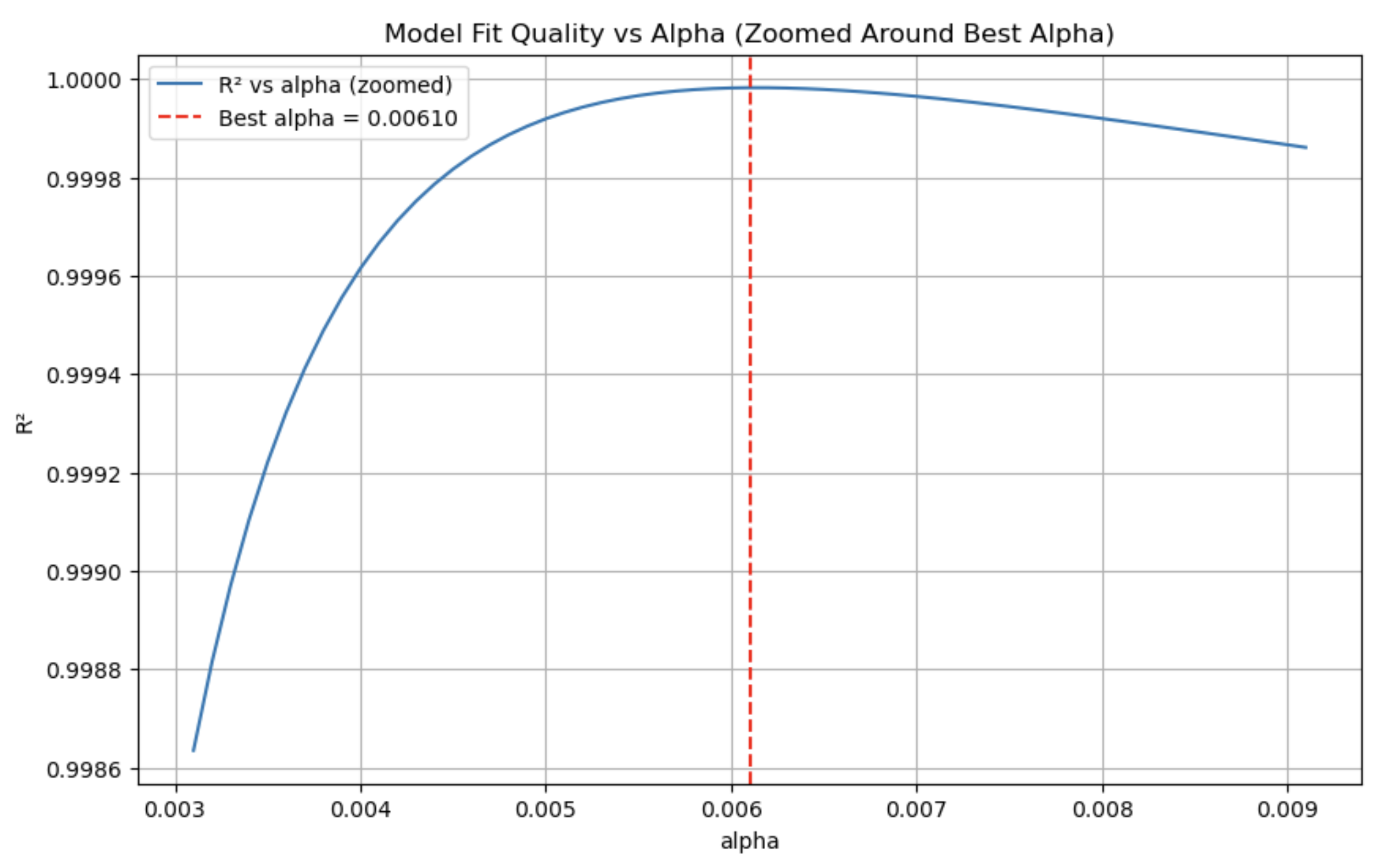}
\caption{The $R^2$ value corresponding to $\alpha = 0.0061$ is the global maximum for the interval $0 \le \alpha \le 1$. For values of $\alpha$ greater than or less than those shown in the graph, the program reports that the $R^2$ values continue declining in either direction.}
\label{RSquared}
\end{figure}

Our code reports that our function best fits the data when $\alpha =$ 0.0061, which aligns with our observations in Desmos. An $\alpha$ of 0.0061 corresponds to an $R^2$ of 0.99998, which indicates an exceptionally good fit. Figure \ref{RSquared} shows the $R^2$ values for different values of $\alpha$ near 0.0061, demonstrating that 0.0061 is indeed the optimal value.

With our value of $\alpha$, we have everything we need to form one continuous position-time function that models the motion of the astronaut for the duration of the entire fall. We will find our final equation in the next section.

\subsection{Final Result}

Let us rewrite Equation \ref{betterswitch} by replacing $x_1(t)$ and $x_2(t)$ with their full equations, as given by Equation \ref{kinematicearly} and Equation \ref{numericalkinematiclate}, and let us replace $\alpha =1$ with $\alpha =0.0061$. Doing this, we acquire:

\begin{equation}
\label{final}
\begin{gathered}
\resizebox{0.93\linewidth}{!}{$ x(t) = \left( - \frac{G M}{2 h^2}t^2 +  v_i t+h \right) \left( \frac{1-\tanh \big({\alpha(t-t_c )}\big)}{2} \right) + \left( - \frac{G M}{2 R^2}t^2 + v_{i2}t + x_{i2} \right) \left( \frac{1+\tanh \big({\alpha(t-t_c )}\big)}{2} \right)$} \\[10pt]
\begin{tabular}{@{\hskip 2pt}l@{\hskip 10pt}l@{\hskip 20pt}l@{\hskip 2pt}l}
$G$   &= $6.6743 \times 10^{-11}\ \text{m}^3\,\text{kg}^{-1}\,\text{s}^{-2}$ &
$M$   &= $5.97219 \times 10^{24}\ \text{kg}$ \\
$h$   &= $1.0000 \times 10^7\ \text{m}$ &
$v_i$ &= $0\ \text{m}\,\text{s}^{-1}$ \\
$\alpha$ &= $0.0061$ &
$t_c$    &= $915$ \\
$R$   &= $6.371 \times 10^6\ \text{m}$ &
$v_{i2}$ &= $5669\ \text{m}\,\text{s}^{-1}$ \\
$x_{i2}$ &= $7.052 \times 10^6\ \text{m .}$ &
         &
\end{tabular}
\end{gathered}
\end{equation}

\begin{figure}[ht]
\centering
\includegraphics[width=0.7\textwidth]{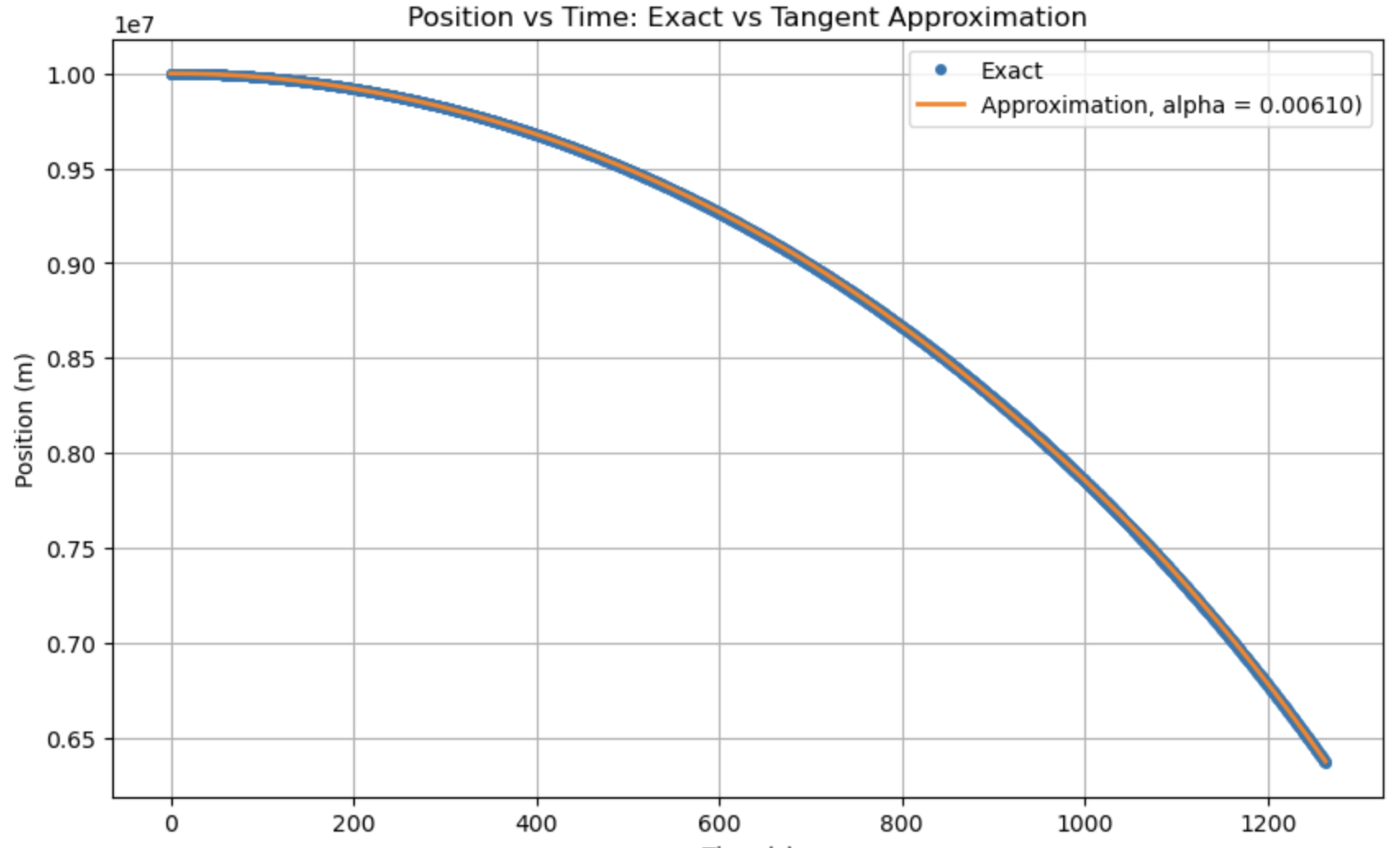}
\caption{The approximate equation $x(t)$ models the exact solution with 99.998$\%$ accuracy.}
\label{GoodFit}
\end{figure}

After pages of calculations, we have finally acquired the equation we have been seeking. Figure \ref{GoodFit} plots Equation \ref{final} alongside the exact value. We see that the approximation is extremely close to the exact values.

Equation \ref{final} provides the position of the astronaut at any moment in time of the students' choosing, which the exact function $t(x)$ was not able to do directly. Furthermore, it is a straightforward analytical equation -- no differential equations or computational methods are needed to understand it.

The pedagogical implications of Equation \ref{final} and the technique it employs serves as the main focus of this paper. In the next section, we explore the educational significance of using the hyperbolic tangent as a tool for modeling systems with time-dependent behavior.

\section{Pedagogical Discussion}

Undergraduate physics courses in classical mechanics often exclude motion with variable acceleration, especially when the course limits itself to analytical solutions. Upper-division courses may explore variable acceleration in the context of differential equations and computational methods, but few introductory courses explore the kinematics of variable acceleration with analytical functions. This gap in introductory physics curriculum is potentially a loss for the student, as variable acceleration is a rich and enlightening area of physics, and it frequently appears in realistic laboratory scenarios. 

The aim of this paper is to present an analytical equation, namely Equation \ref{final}, that is approachable to undergraduate students but captures the complexities of variable acceleration. In other words, Equation \ref{final} is proof by example that computational methods are not always essential for modeling motion in complex situations. This method of using switch functions therefore makes a branch of physics problems accessible to students who struggle with computational methods or numerical approximations.

The switch-function method also presents undergraduate students with a novel lens through which to view functions and operators. In the technique presented above, the equation for acceleration remains time-independent, and time dependency was applied not by changing the acceleration variable itself, but by changing the whole equation of motion. In this way, the hyperbolic tangent acts more akin to an operator than a function -- it changed the nature of an entire function rather than an individual variable. Introducing students to operators early in their physics curriculum has the potential to strengthen their conceptual foundations for higher-level topics. In particular, quantum mechanics frequently utilizes operators, especially the time dependency operator. While quantum mechanics can sometimes be highly abstract and difficult to visualize, the Falling Astronaut Problem presents an opportunity for students to conceptualize operators in a concrete scenario that is easier to picture.

The switch-function technique has broad applications beyond the physics classroom. In the context of aerospace engineering, for example, rocket thrusters may be turned on and off during different time intervals, and switch-functions provide an elegant way to model this behavior. Indeed, any situation in which there is an abrupt change in the circumstances at known moments in time could potentially be modeled with the switch-function technique. Situations with abrupt but predictable changes appear frequently in biology, meteorology, and economics.

The switch-function technique can also be generalized to the ``smoothing-function" technique. When the rate of transition $\alpha$ is much less than 1, as in the Falling Astronaut Problem, the switch-functions begin to behave like dials that smoothly transition a system one state to another. In such a situation, the term ``smoothing function" is more appropriate than ``switch function." 

The smoothing-function technique can be utilized to create analytical solutions for a huge diversity of situations that may otherwise require differential equations and computational methods. Examples include toy magnets attracting each other on a tabletop or a rocket with constant thrust that is steadily losing mass as fuel is ejected. Beyond the physics classroom, smoothing functions can help model stock prices, weather patterns, and plant growth. Future research into smoothing functions may reveal approximate analytical equations in situations that would be otherwise inaccessible to undergraduate students.


\section{Acknowledgments}

Thank you to Elizabeth Cavicchi for her mentorship and guidance during the publication process of this paper and its predecessor.

Thank you to Laura Fledderman for her encouragement and feedback during the writing of this paper.

\end{document}